# Title: Scaling Law of Urban Ride Sharing

**Authors:** R. Tachet[1], O. Sagarra[1,2], P. Santi[1,3*], G. Resta[3], M. Szell[1,4], S. H. Strogatz[5], C. Ratti[1]

**Affiliations:**

[1] *Senseable City Lab, Massachusetts Institute of Technology, Cambridge, MA 02139, USA.*

[2] *Complexity Lab Barcelona, Universitat de Barcelona, 08028 Barcelona, SPAIN.*

[3] *Istituto di Informatica e Telematica del CNR, 56124 Pisa, ITALY.*

[4] *Center for Complex Network Research, Northeastern University, Boston, MA 02115, USA.*

[5] *Department of Mathematics, Cornell University, Ithaca, NY 14853, USA.*

[*] Correspondence to: rtachet@mit.edu

**Abstract:**

Sharing rides could drastically improve the efficiency of car and taxi transportation. Unleashing such potential, however, requires understanding how urban parameters affect the fraction of individual trips that can be shared, a quantity that we call *shareability*. Using data on millions of taxi trips in New York City, San Francisco, Singapore, and Vienna, we compute the shareability curves for each city, and find that a natural rescaling collapses them onto a single, universal curve. We explain this scaling law theoretically with a simple model that predicts the potential for ride sharing in any city, using a few basic urban quantities and no adjustable parameters. Accurate extrapolations of this type will help planners, transportation companies, and society at large to shape a sustainable path for urban growth.

**Main text:**

Mobility of people and goods has been vital to urban life since cities emerged more than 7,000 years ago [*1*]. Indeed, the success, prosperity, and livability of cities are directly related to the effectiveness of their mobility systems [*2*]. However, due to fixed schedules, limited coverage, and low quality of travel experience, public transportation systems accommodate only a fraction of the urban mobility demand [*3*]. The rest is satisfied by private vehicles and taxis, inefficient transportation modes that move only 1.3 passengers per vehicle on average [*4*, *5*], causing the road congestion observed in most cities worldwide, with immense economic and societal costs [*6*]. Enhancing transportation efficiency is a key to rendering sustainable the urban growth predicted for the coming years [*7*].

The emerging sharing economy [*8*, *9*] promises to improve the efficiency of individual, on-demand transportation. Bridging the gap between shared but inflexible public transportation and flexible but not shared private transportation, novel services such as those provided by Uber™,

Lyft™, and ZipCar™ can significantly contribute to reducing road congestion and emissions. But the realizability of such potential benefits depends on the answer to a fundamental unsolved question: How compatible in space and time – and thus shareable – are individual mobility patterns?

While recent literature [*10-15*] has unveiled spatial and temporal regularity of individual mobility patterns, very little is known about their mutual similarity. In a previous study [*16*], we introduced the notion of a shareability network to quantify the spatial and temporal compatibility of individual trips. The nodes in the network represent trips, and links between them mark trips that can be shared. Two trips are defined to be shareable if they would incur a sharing delay of no more than $\Delta$ minutes, relative to a single ride (see Supplementary Information). Let the shareability metric $S$ denote the fraction of individual rides that can be shared. We found [*16*] that taxi trips in New York City offer a shareability well above 95% for $\Delta = 5$ min, and that $S$ increases rapidly with the number of trips available for sharing.

But that previous study [*16*] left a key question unresolved: Might the results be peculiar to New York City? There was good reason to suspect so, given that New York is singular in several respects, namely, its large population, its small geographical area, and its enormous density of taxi traffic. In what follows, we show that three other major world cities, which differ greatly from each other and from New York City in their traffic characteristics, population size, and geographical area, all obey the same empirical law governing the potential for ridesharing. To the best of our knowledge, the existence of such a seemingly (we only have four cities at our disposal) universal law has not been reported before. We explain the mechanism underlying this law of ride sharing using a simple mathematical model. The model's prediction accounts for more than 90% of the variance in the data, and does so without any adjustable parameters. What is important here is the generality of the law, as well as its rapidly saturating shape, because together they imply that ride sharing could have a large beneficial impact in virtually any city, not just New York City.

**Results**

Let *C* be a city, *Ω(C)* its spatial domain, |*Ω(C)*| its area, *v(C)* the average traffic speed in *C* and *λ* the average number of trips per hour with both endpoints in *Ω(C)*. Figure 1(a) shows that the computed curve of shareability against *λ* for New York City [*16*] closely resembles a "fast" saturation process, with a quick increase from lowest density, where shareability is minimal, to saturation where all trips can be shared. Our first main finding is that three other cities – San Francisco, Singapore, and Vienna (see Methods and Supplementary Information: Table S1a for datasets description and algorithms [*17*]) – show strikingly similar shareability curves (Figures 1(b)-(d)). Such a similarity is remarkable, given that the shareability curves are obtained from data sets of real taxi trips, using a methodology that includes the hour-by-hour variability in traffic congestion (see Methods).

Each curve in Figure 1 saturates rapidly as a function of *λ*. Their rapid saturation distinguishes them from other saturation phenomena observed in urban/geographical processes, such as the growth of retail locations [*18*] and the spreading of innovations [*19*], which are instead characterized by an initial "slow start" phase with a sigmoidal shape. Fast saturation of shareability is a plausible explanation for the great success of innovative ride and vehicle sharing apps such as UberPool™, ZipCar™, and Car2Go™.

The similarity we observe between cities actually goes beyond the resemblance of their shareability curves: a single linear rescaling of the *λ*-axis makes all the curves nearly coincident (Figure 2; see also Methods and Supplementary Information: Table S2), suggesting that a common mechanism governs shareability in those four cities. The data collapse is achieved by replotting the computed shareability $S$ versus the dimensionless quantity

$$L = \lambda \Delta^3 \frac{v^2(C)}{|\Omega(C)|}, \tag{1}$$

The greater the $L$, the greater the shareability.

The quality of the data collapse indicates that a few urban level parameters, when combined into the dimensionless group $L$, suffice to accurately model a complex quantity like the fraction of trips that can be shared in a city. This result is all the more surprising when one considers that $L$ is defined in terms of the *average* daily traffic speed in the city, while the shareability curves have been derived using hourly, street-level traffic speed estimations. The implication is that the effect of traffic congestion on shareability is limited.

The particular combination of urban parameters in Eq. (1) can be rationalized by dimensional analysis. Intuitively, $L$ represents a ratio between two timescales: the sharing delay $\Delta$ and the characteristic waiting time $t_{\text{wait}}$ for a trip to be generated in a user's vicinity. To see this, imagine that you are looking for a cab. Since $\lambda$ is the average rate at which taxi trips are generated, $1/\lambda$ is the characteristic time for a new trip to be generated, somewhere in the city. But the city as a whole is not what concerns you. What matters more is how long you can expect to wait for a new trip to be generated in your vicinity. The characteristic linear scale of a vicinity is $v\Delta$, the distance a cab moving at speed $v$ would travel in the delay time $\Delta$ that another passenger could tolerate. Since the city has a total area $\Omega$ and each vicinity has area $(v\Delta)^2$, there are about $|\Omega|/(v\Delta)^2$ vicinities in total. Assuming that trips are generated uniformly in space, you would expect a trip to be generated in your vicinity every $t_{\text{wait}} = \frac{1}{\lambda} \times \frac{|\Omega|}{(v\delta)^2}$ time units. Hence the ratio of the tolerable delay time $\Delta$ to the expected waiting time is $\Delta / t_{\text{wait}} = \lambda v^2 \Delta^3 / |\Omega| = L$.

At a more refined level, the influence of urban parameters on shareability can be approached mathematically as follows. Intuitively, one expects that shareability should be positively related to $\Delta$, $v(C)$, and $\lambda$. Indeed, as $\Delta$ and $v(C)$ increase, people become more tolerant about sharing delay and a larger urban space can be covered without exceeding the delay [16]. The effect of increasing trip density is more complex to assess since it simultaneously introduces new rides and new ride-sharing opportunities. However, the additional trips are drawn from the same distribution as the original ones, so they possess similar spatiotemporal properties, which on average results in an increase of shareability as a function of $\lambda$.

Assuming that rides are generated independently in the city according to a given spatiotemporal distribution, we wish to compute the probability that a ride can be shared as a function of $\Delta$, $v(C)$, and $\lambda$. Tackling this problem directly is very difficult, since the probability of actually sharing a ride depends not only on the spatiotemporal availability of candidate trips to share, but also on how potentially shareable trips are paired together, which in turn depends on complex structural properties of the underlying shareability network. Nevertheless, the spatial dimension of the problem, coupled with the observed fast saturation of the shareability curve, suggest analogies with geometric random graphs [20] and percolation theory [21]. A common trait of these theories is that complex network structural properties such as connectivity can be closely approximated by much simpler properties, such as the existence of isolated nodes. This turns out to be the case also for shareability networks; we find that shareability $S$ is highly correlated with the number of isolated nodes in the shareability network (Methods).

Based on the above discussion, we can model shareability by fixing an arbitrary trip $T$ and estimating the probability that there exists at least one other trip $T'$ shareable with $T$. More specifically, an arbitrary trip $T$ starting at time $t_0$ and going from origin $o$ to destination $d$ defines a trajectory in space and time. For fixed $\Delta$ and average traffic speed $v(C)$, we define the notion of the shareability shadow $s(T)$ surrounding $T$ and confining the region of sharing opportunities (Supplementary Information: Figs. S1 and S2). For another trip $T'$ to be shareable with $T$, its trajectory needs to overlap (i.e., to take place at the same time, at least partially) and to be "aligned" (i.e., not deviate too much direction-wise) with $s(T)$. Those two conditions simply translate our upper bound $\Delta$ on delays into a geometric condition stating that shareable trips should be close enough in terms of trajectories, where close enough is quantified through the volume of $s(T)$ chosen depending on $v(C)$ and $\Delta$. Analytically, the expected shareability becomes the probability that a compatible trip will be generated in the shareability shadow (see Supplementary Information: "Supplementary Equations"). To compute that quantity, the previously mentioned spatiotemporal distribution of trips has to be determined. Among the different options we considered, the following one gave the best compromise between accuracy and tractability: origin point $o$ chosen uniformly in $\Omega$, and destination point $d$ chosen uniformly in a disk centered on $o$ of radius $R$ (ignoring boundary effects for the sake of simplicity). The geometry of the city plays a minimal part in the definition, which allows us to derive analytical formulas for the shareability. For $R$ large enough, we find that $S$ becomes independent of $R$, and the city's influence on the shareability only appears through the quantity $L$. We prove that (see Supplementary Information: "Supplementary Equations")

$$S = 1 - \frac{1}{2L^3}(1 - e^{-L})(1 - (1 + 2L)e^{-2L}). \tag{2}$$

We tested our model predictions on the four cities mentioned above and found a strong agreement with the respective shareability curves (Supplementary Information: Fig. S3), with $R^2$ values ranging from 0.91 to 0.98.

**Discussion**

The fidelity of the model suggests that the fundamental mechanisms governing ride sharing in a real world scenario where trips are performed in a road network with traffic congestion can be accurately characterized through a relatively simple mathematical model built upon a number of simplifying assumptions: Euclidean geometry, straight-line trajectories, and extremely basic shareability shadow shapes. In particular, most of the knowledge required to determine shareability is contained in the dimensionless group $L$. Being relatively easy to estimate, and the only quantity required for our (otherwise parameter-free) framework, $L$ gives the model strong predictive power.

A final important feature of the framework is its flexibility, which allows for potential enhancements. For instance, if the impact of congestion on average speed were known, through modeling or pervasive sensors, the model could be extended to take second-order effects into account (ride sharing reduces congestion and increases average travel speed, thereby increasing shareability; see Supplementary Information: "Supplementary Equations").

Our findings quantify the effects of ride sharing on the urban environment and shed light on the recent upheaval that ride sharing has caused in cities worldwide. Furthermore, they offer valuable guidance towards designing more efficient mobility systems in the future. Tables 1 and 2 show the urban parameters and corresponding predictions for ride shareability in several major world cities. Even for low trip density, and allowing delays no longer than $\Delta = 5$ minutes, the potential for sharing is massive.

**Methods**

The New York dataset has been obtained from the New York Taxi and Limousine Commission for the year 2011 via a Freedom of Information Act request. It is the same as the dataset used in [16, 17]. The San Francisco dataset is freely available [24]. The Vienna and Singapore datasets were provided to the MIT SENSEable City Lab by AIT and the Singapore government, respectively.

The New York dataset spans more than an entire year and contains all taxi trips generated in the area of New York by its approximately 13,500 taxis. The other datasets span roughly over a month and contain records provided by a single taxi operator. The total number of cabs in San Francisco is officially 1494 [25], and the number of taxis tracked in the data set is about 500. For Singapore, the official figure is 25176 [26], and our data set refers to about 16,000 taxis. For Vienna, we have traces of about 1,000 taxis, while the total number of taxis operating in the city is unknown. Data set details are reported in the Supplementary Information, Table S1.

We have applied the same filtering procedure to all the datasets: only trips performed while a customer occupied the taxi were considered in the analysis. From these trips, we only kept the ones with start and end GPS positions within 200 meters of the closest intersection present in the considered area of study. Such an area was obtained by considering the borough of Manhattan (NY), the entire island of Singapore (SI), and both the urban areas of San Francisco (SF) and Vienna (VI), including the road to the airport. We included the airports of San Francisco, Singapore, and Vienna since trips to and from them account for a substantial fraction of the dataset.

The intersections were obtained from Open Street Map [27] (see Supplementary Information, Fig. S4), considering only primary and secondary level roads and by manually merging all repeated elements corresponding to every given intersection (using GQIS [28]). All trip coordinates were provided in longitude-latitude pairs using the *WGS84* ellipsoid but have been projected to Euclidean UTM coordinates using the zones specified in the Supplementary Information, Table S1a.

After pre-processing, each trip is uniquely identified by a tuple containing starting and ending (*latitude*, *longitude*) coordinates, which correspond to the coordinates of the intersections closest to start and ending coordinates of a trip, and by a pickup and dropoff time. Pickup and dropoff times are used to estimate travel times between any two intersections in the city for each of the 24 hours, according to the procedure described in [16]. This method allows accounting for the effect of traffic on travel time when computing the shareability networks used to obtain the shareability curves shown in the paper. Shareability networks for the four cities were obtained using the method described in [16].

To generate the saturation curves used in the paper, two procedures were required. For the New York, San Francisco and Singapore datasets, for which the shareability curves are saturated, the lower parts of the curves (corresponding to small trip densities $\lambda$) were obtained by randomly and uniformly subsampling the database of actual trips up to the desired density. For the Vienna case, a second procedure was necessary to reach densities higher than those in the dataset (explaining why Vienna curves show $\lambda$ and $L$ values larger than the $\lambda_f$ and $L(C)$ from Table S1b reported in the Supplementary Information). We call that procedure *supersampling*, and extend an existing method [17].

The above procedure is used to interpolate trips from a given sample in a static manner in time. It is based on inferring a city's invariant collection of transition probabilities $\{p_{ij}\}$, where $ij$ enumerates all possible intersection pairs. Such a collection of values is normalized ($\Sigma_{ij} p_{ij} = 1$) and represents the probability that a given trip is generated at intersection $i$ and ends at intersection $j$. Such a collection is shown [17] to be extremely stable in time, and a procedure is developed to infer the complete set of values (note that in general $p_{ij} \neq 0$ for all $i$ and $j$). Once this collection of values is obtained, for a given density (total number of trips $T_{trips}$ generated in a given timespan $\tau$) the allocation of trips to each intersection pair $ij$ reads $<t_{ij}> = T_{trips} \, p_{ij}$ with $t_{ij}$ an integer random variable following a Poisson distribution.

We have extended the method of [7] to allow for *dynamic supersampling* in time. Algorithm 1 (see Supplementary Information, Algorithm S1) exploits the exponential nature of inter-events times between trips (see Supplementary Information, Fig. S5) coupled with the statistics of daily and hourly trip generation (see Supplementary Information, Fig. S6). For every day, the algorithm distributes the empirical number of daily generated trips $T_d$ over hourly intervals according to the empirical probability $q_h = \check{T}_h / \Sigma_{\hat{h}} \check{T}_{\hat{h}}$ (where $\check{T}_h$ is the average number of trips observed during hour $h$ and $0 \leq h, \hat{h} \leq 23$) and then distributes the generated trips over intersections according to $p_{ij}$. Finally, for each intersection, the number of allocated trips is distributed in time according to a Poisson process. The code for this extension was made public [29].

**Acknowledgements:**

The datasets used for this study have been obtained from different sources. The New York taxi data set is publicly available, and has been obtained directly from the New York Taxi and Limousine Commission. The San Francisco taxi data set is publicly available at the following URL: http://crawdad.org/epfl/mobility/20090224/. The Vienna taxi data set has been obtained by AIT, while the Singapore taxi data set has been obtained from the Singaporean government. Sample of the two latter data sets will be made available upon request.

R. Tachet, P. Santi, and C. Ratti thank ENEL Foundation, Accenture China, American Air Liquide, Emirates Integrated Telecommunications Company , Ericsson, Kuwait-MIT Center for Natural Resources and the Environment, Liberty Mutual Institute, Singapore-MIT Alliance for Research and Technology, Regional Municipality of Wood Buffalo, Volkswagen Electronics Research Lab, and all the members of the MIT Senseable City Lab Consortium for supporting this research. Research of S.H.S. was supported by NSF Grants DMS-1513179 and CCF-1522054.

**Author contributions:**

R.T. analyzed the data, worked on the mathematical model, and contributed to writing. O.S. processed and analyzed the data, worked on the mathematical model, and contributed to writing. P.S. designed the research and trip matching algorithms, assisted in the mathematical model, and contributed to writing. G.R. built the shareability networks, design and executed the trip matching algorithms. M.S. processed the data. S.H.S contributed to the dimensional analysis and writing. C.R. supervised the research and contributed to writing.

**Competing Financial Interests:**

The authors claim no competing financial interest.

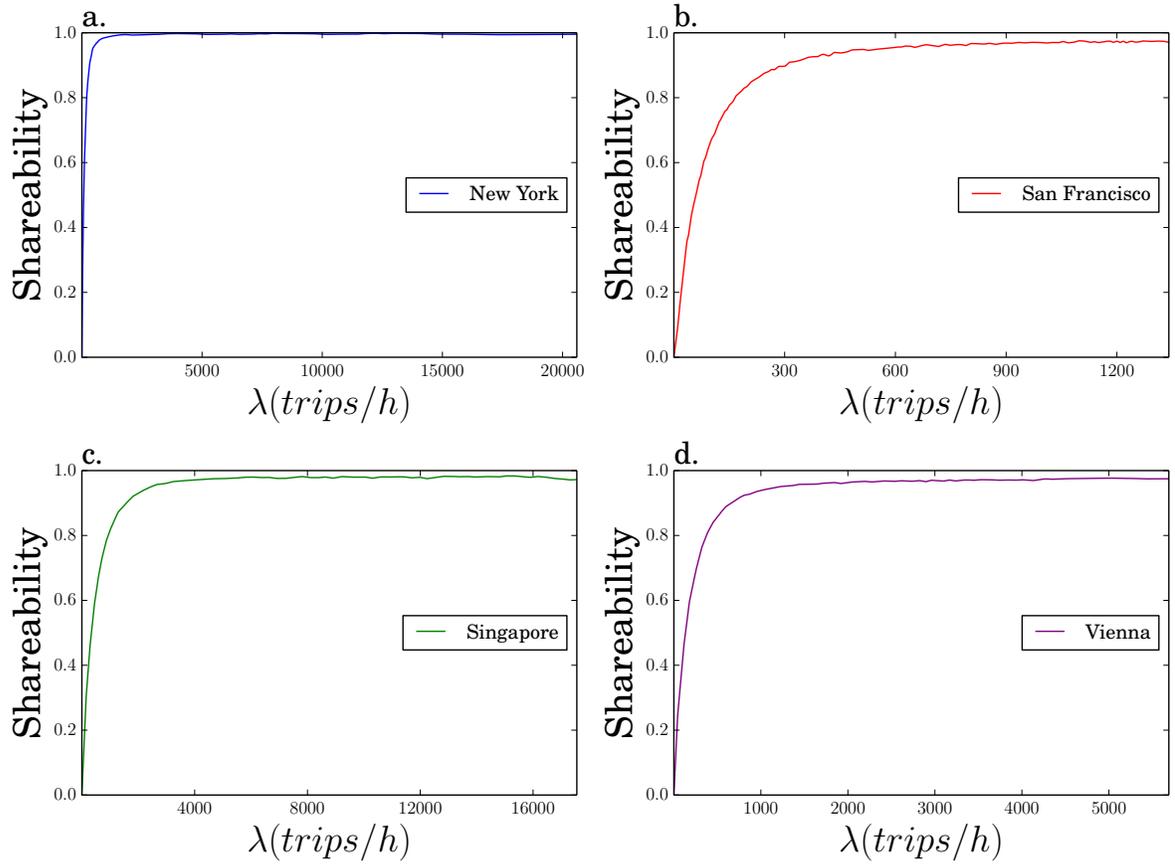

**Figure 1**. **Shareability curves.** The shareability curves for (a) New York, (b) San Francisco, (c) Singapore, and (d) Vienna. The curves were computed using a shareability network algorithm (*16*) applied to data collected from over 156 million taxi trips in the four cities. See Methods for details about the datasets and algorithm.

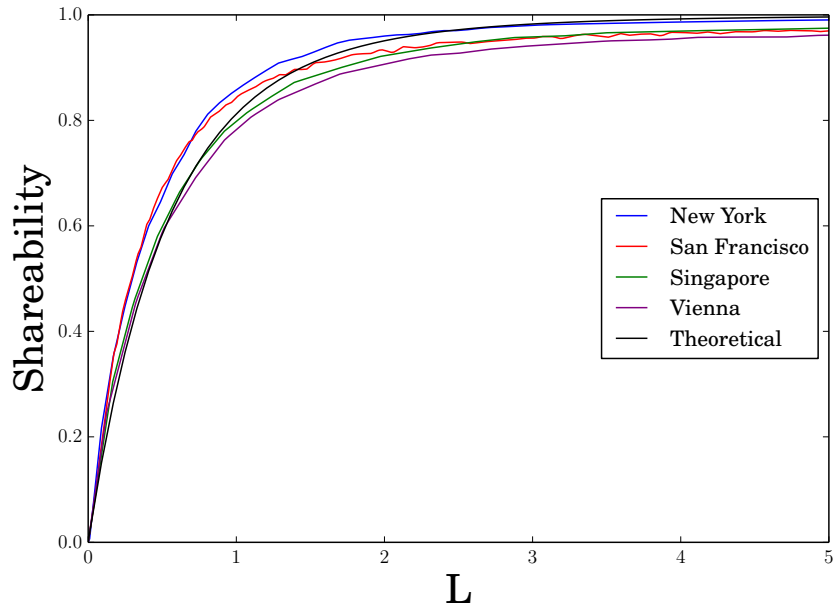

**Figure 2**. **Shareability law.** When re-plotted as functions of *L*, the computed shareability curves for New York, San Francisco, Singapore, and Vienna nearly coincide with each other and with the theoretical prediction given by Eq. (2). This rescaling involves no adjustable parameters.

**City Parameters**

| Cities | Amsterdam | Berlin | London | Newcastle | Paris | Prague | Rome | Santiago |
|---|---|---|---|---|---|---|---|---|
| Area ($km^2$) | 219 | 892 | 1572 | 360 | 105 | 496 | 1285 | 641 |
| Speed ($km/h$) | 34 | 19 | 19 | 42 | 31 | 37 | 30 | 31 |

**Table 1. City parameters.** Area ($km^2$) and average speed ($km/h$) of vehicles in different cities around the world (*22*, *23*).

**Shareability**

| | Cities | Amsterdam | Berlin | London | Newcastle | Paris | Prague | Rome | Santiago |
|---|---|---|---|---|---|---|---|---|---|
| *Trips / h / km²* | 0.5 | 45% | 17% | 17% | 59% | 39% | 51% | 37% | 39% |
| | 2.5 | 92% | 60% | 60% | 97% | 89% | 95% | 88% | 89% |
| | 4.5 | 98% | 80% | 80% | 99% | 97% | 99% | 97% | 97% |
| | 6.5 | 99% | 89% | 89% | 100% | 99% | 100% | 99% | 99% |
| | 8.5 | 100% | 94% | 94% | 100% | 100% | 100% | 99% | 100% |

**Table 2**. **Shareability in different world cities.** Shareability as a function of spatiotemporal trip density, measured in units of trips per hour per square kilometer. For comparison, the spatiotemporal trip densities of taxi rides in our datasets are: 344.12 *trips/h/km²* in New York, 24.46 in Singapore, 12.63 in San Francisco and 0.95 in Vienna (Supplementary Information Table S1b), generating a shareability of almost 100% for the first three and of 83% for Vienna.

# Supplementary Information for

## Scaling Law of Urban Ride Sharing

R. Tachet, O. Sagarra, P. Santi, G. Resta, M. Szell, S. H. Strogatz, C. Ratti

correspondence to: rtachet@mit.edu

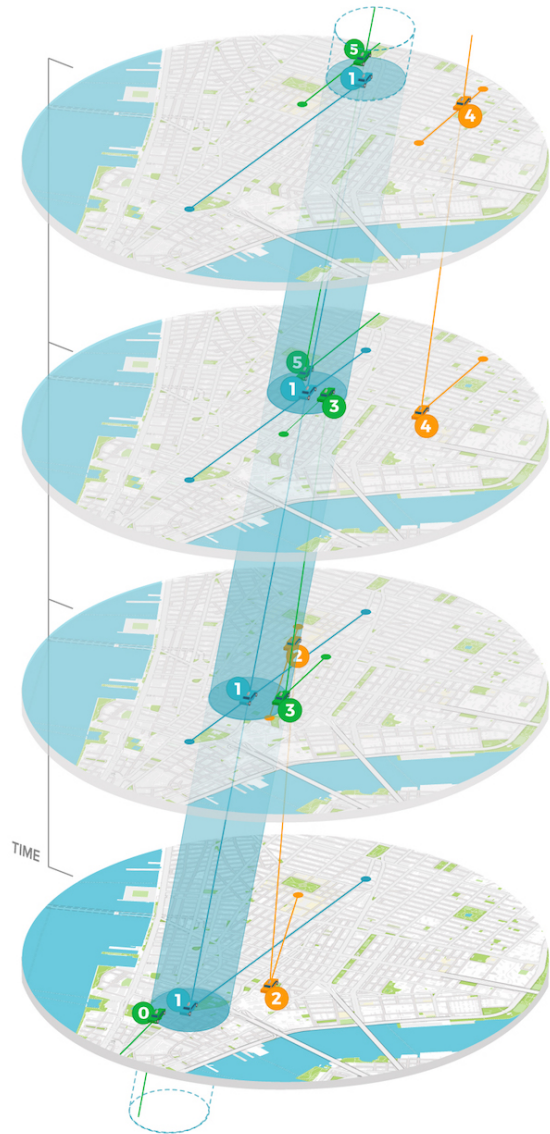

**Fig. S1. Shareability shadow.** Two-dimensional representation of city *C*, extruded to highlight the temporal dimension. The blue segment represents the trajectory *T* of a given trip. The extended cylinder surrounding it represents its shareability shadow *s(T)*. For a trip to be shareable, its endpoints must belong to *s(T)* (which is the case for trips 0, 3 and 5), and at least one of them must belong to the darker part of the cylinder (only 3 and 5 satisfy this condition). Trip 2 is spatially compatible with *T*, but not temporally (it started too early). Trip 4 is neither spatially nor temporally compatible with *T*. This map was generated using MapBox and Adobe Photoshop CC (https://www.mapbox.com/ and http://www.adobe.com/products/photoshop.html).

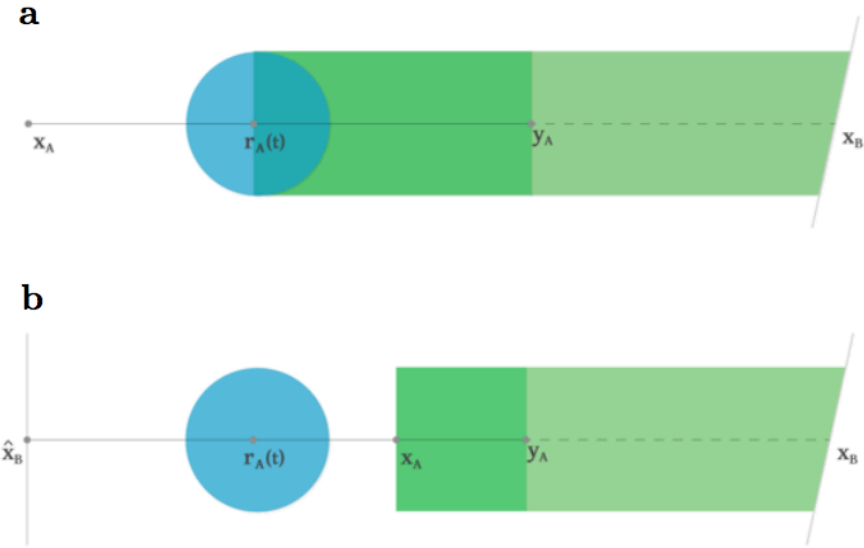

**Fig. S2. Projected shareability shadows. a**, *Forward sharing probability.* Schema representing the different elements involved in the integration of $\mathcal{A}$ (t | $t_A$, $x_A$, $y_A$). $\varepsilon^{\Delta}(r_A(t))$, $\Xi(r_A(t), y_A)$, and $\Xi(y_A, x_B)$ correspond respectively to the blue disk, the dark green rectangle, and the light green quadrilateral. **b,** *Backward sharing probability.* Schema representing the different elements involved in the backward integration of $\mathcal{A}$' (t | $t_A$, $x_A$, $y_A$). Similarly, $\varepsilon^{\Delta}(r_A(t))$, $\Xi(r_A(t), y_A)$, and $\Xi(y_A, x_B)$ correspond respectively to the blue disk, the dark green rectangle, and the light green quadrilateral.

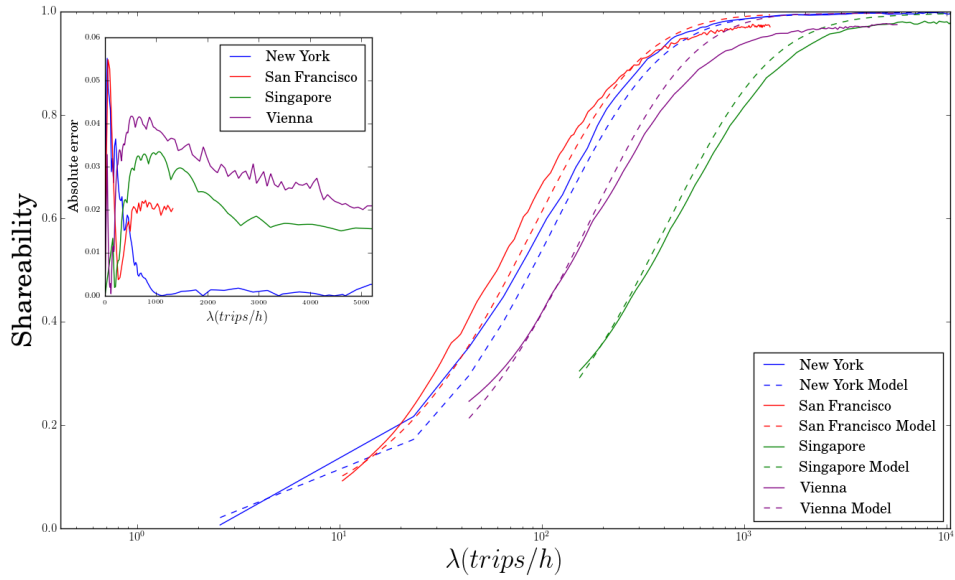

**Fig. S3. Model accuracy.** This log-lin plot assesses the accuracy of our mathematical model for shareability against real data for New York, San Francisco, Singapore and Vienna. The model provides an accurate estimate of the shareability in those cities, with $R^2$ values respectively equal to 98.9%, 97.7%, 95.0% and 91.4%. Subplot: *Errors*. Absolute errors between model predictions and real data for the four cities considered in our study.

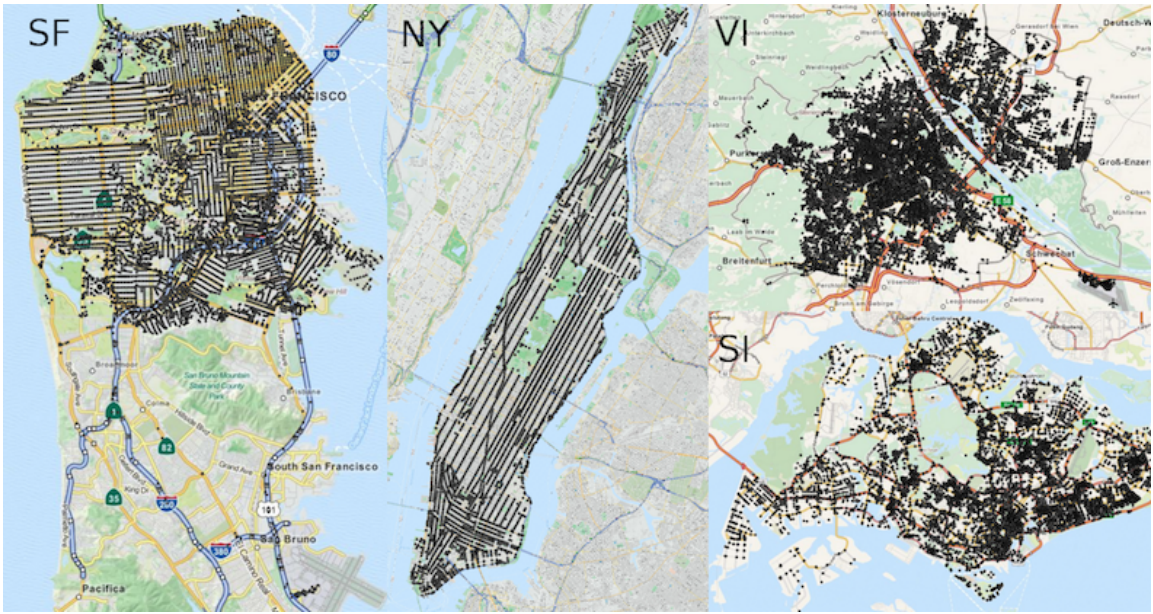

**Fig. S4. Maps of the intersections used for trip filtering.** Maps of the four different cities (SF – San Francisco, NY – New York, VI – Vienna, and SI – Singapore) for which we have data on taxi displacements with the considered intersections for trip matching overprinted using black dots. The San Francisco, Singapore and Vienna cases include the airport as it concentrates a relevant fraction of the total traffic. Background maps obtained from Open Street Map (27, http://www.openstreetmap.org/copyright).

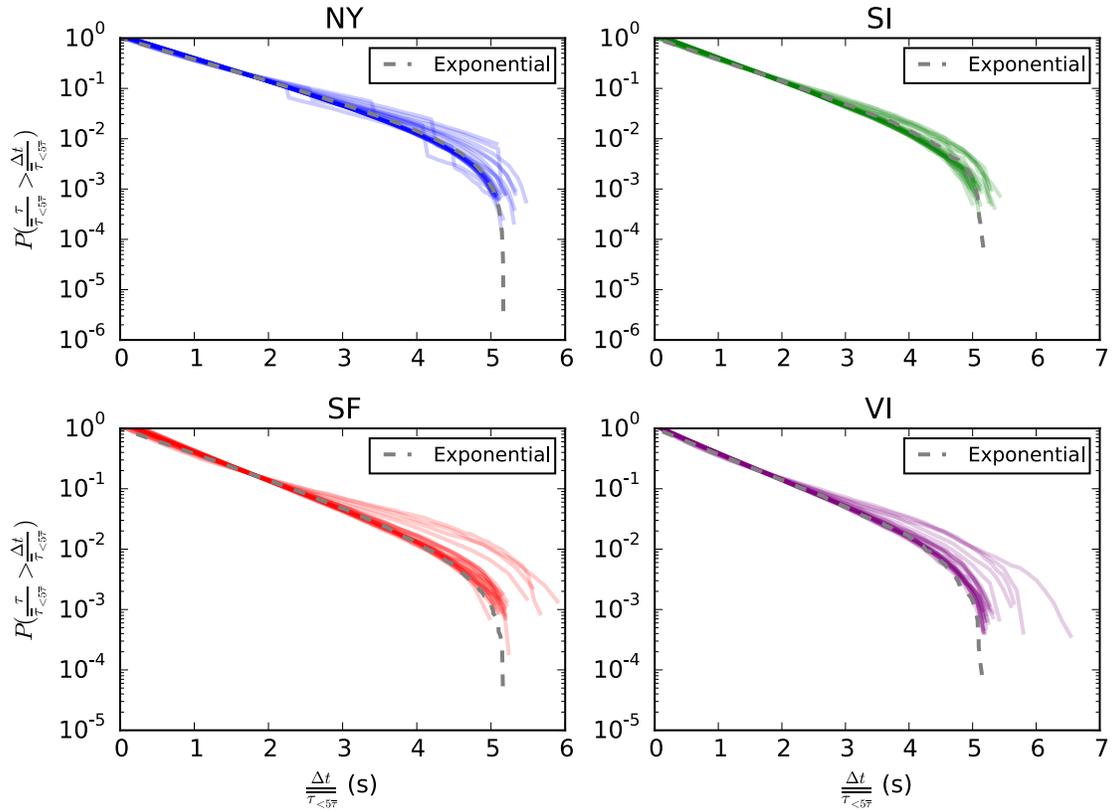

**Fig. S5. Evidence of Poisson trip generation.** Hourly inter-event time distribution compared to an exponential distribution (which would correspond to purely Poisson distributed trips). Each line corresponds to a different hour of day and transparency has been applied to visualize the density of curves.

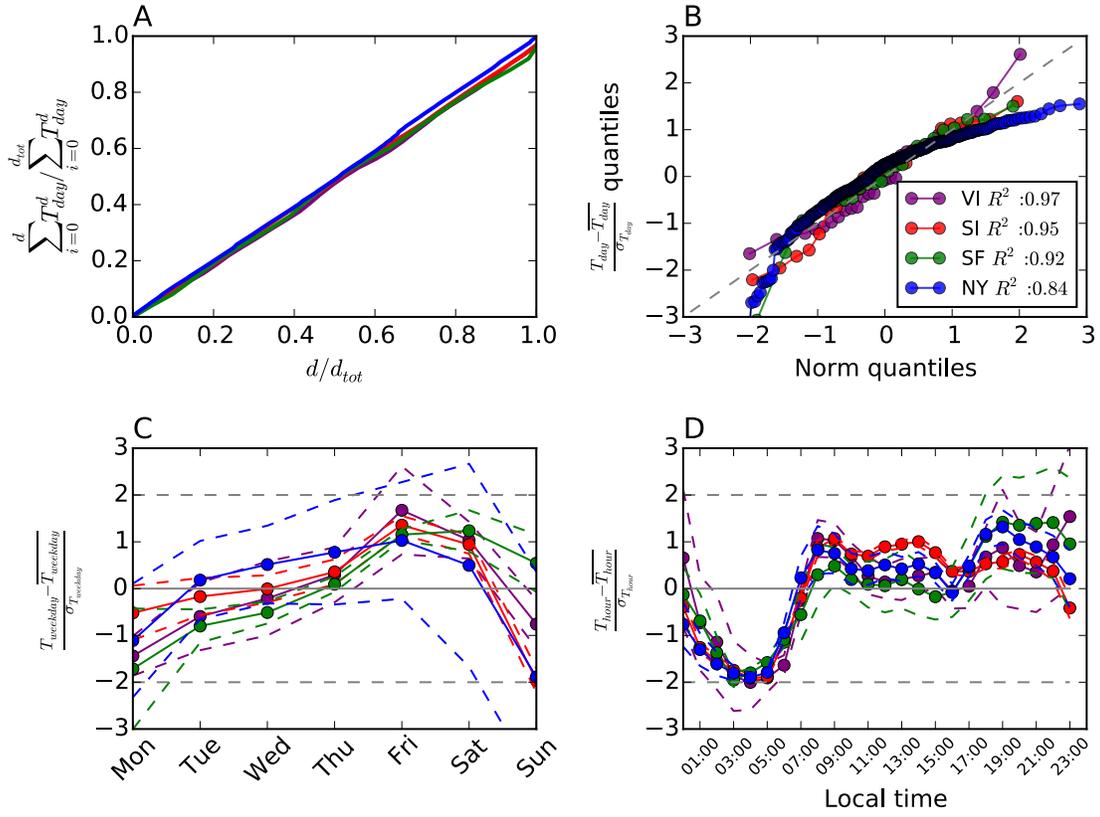

**Fig. S6. Trip generation.** A, Evolution of accumulated number of recorded trips with time. B, Quantile-quantile plot of its distribution compared to a Normal curve with $R^2$ of the linear fit also shown. C~~, Standardized average daily. D, hourly time generation of trips for the datasets also displayed.

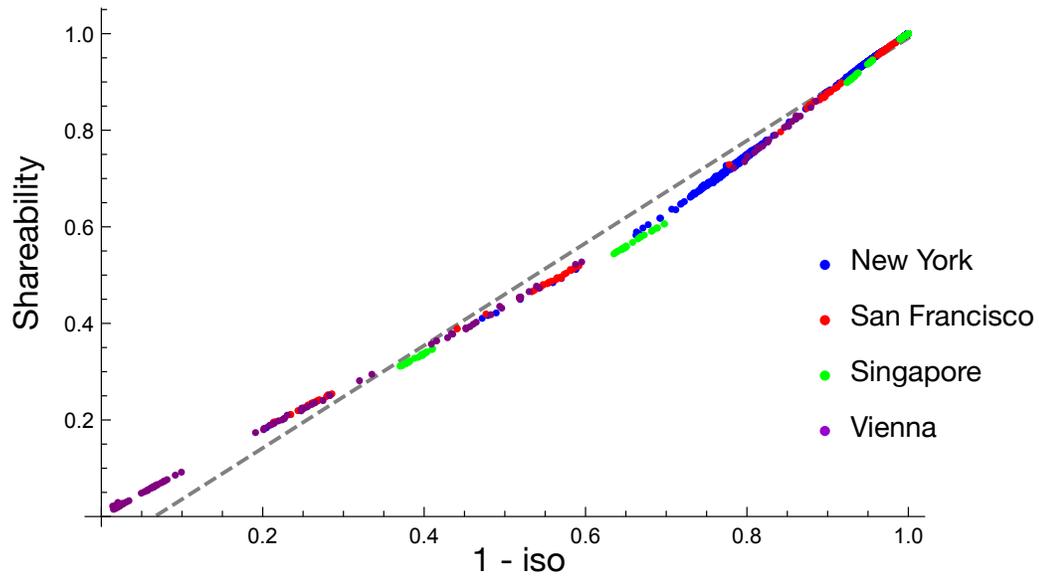

**Fig. S7. Correlation between shareability and isolated nodes.** Shareability can be accurately predicted using the fraction of isolated nodes in the shareability network. The plot shows the very good statistical correlation between fraction of isolated nodes (*x* axis) and shareability (*y* axis), with $R^2$=0.99. Data from the New York, San Francisco, Singapore, and Vienna data sets.

**a**

| Dataset | $N_{taxis}$ | $T$ | $\rho_{taxis}$ | UTM Zone | Dates |
|---------|-------------|-----|----------------|----------|-------|
| New York | 13052 | 146986835 | 1 | 18 | 1/01/2011-31/12/2011 (365 days) |
| San Francisco | 537 | 435670 | 0.35 | 10 | 17/05/2008-10/06/2008 (24 days) |
| Singapore | 15915 | 8873029 | 0.6 | 48 | 14/02/2011-13/03/2011 (27 days) |
| Vienna | - | 284541 | - | 33 | 28/02/2011-31/03/2011 (31 days) |

**b**

| | $v(C)$ | $|\Omega(C)|$ | $\lambda_f$ | $\rho$ | $L(C)$ |
|---|--------|---------------|-------------|--------|--------|
| New York | 20.1 | 59.9 | 20612.7 | 344.12 | 80.46 |
| San Francisco | 32.0 | 121.4 | 1342.0 | 24.46 | 6.55 |
| Singapore | 36.7 | 718.3 | 17569.3 | 12.63 | 19.07 |
| Vienna | 44.3 | 395.3 | 395.5 | 0.95 | 1.08 |

**Table S1. Taxi trip datasets and city characteristics. a**, $N_{taxis}$ refers to the number of taxis present in the dataset and $\rho_{taxis}$ to the fraction of the entire taxi fleet it represents. $T$ refers to the total number of recorded trips. Used UTM zones for projection in trip intersection matching are also shown. **b**, Average speed of vehicles (in *km/h*), area (in *km²*), average number (in *trips/h*) and spatiotemporal density (in *trips/h/km²*) of taxi trips, and $L$ (dimensionless) for New York, San Francisco, Singapore and Vienna, with $\Delta = 1/12$ h.

|       | Opt. $R^2$ | $L(C)$. $R^2$ |
|-------|------------|---------------|
| NY-SF | 0.981      | 0.978         |
| NY-SI | 0.972      | 0.911         |
| NY-VI | 0.955      | 0.856         |
| SF-SI | 0.994      | 0.956         |
| SF-VI | 0.998      | 0.930         |
| VI-SI | 0.991      | 0.984         |

**Table S2. Similarity measures between the different curves.** *Opt. $R^2$* was obtained using the optimal rescaling, and *$L(C)$. $R^2$* using *$L(C)$*. In both cases, the values are large, confirming the visual impression of a strong agreement between the rescaled curves.

---
**Input:** Number of intersections $N$, City intersection pair invariants $\{p_{ij} \forall ij = 1...N^2\}$ (float vector), Number of trips per day $\{T_d, \forall d = 0...d_{days}\}$ (float vector), Hourly probability of trip allocation $\{q_h \forall h = 1...24\}$ (float vector), number of days to sample $d_{days}$ (int), inflation factor $f$ and starting time $\tau_0$. .
**Output:** List of generated trips $(i, j, \tau) \forall t = 1...T$ (vector with each entry a 3 float tuple).
**begin** Initialization
$\quad$ Set $d_{type} = 0$. Set $d = 0$. Set $\vec{L} = 0$. Set $\tau = \tau_0$;
**end**
**begin** Day generation
$\quad$ **while** $d < d_{days}$ **do**
$\quad\quad$ **begin** Hourly generation
$\quad\quad\quad$ **for** $h = 0, 23$ **do**
$\quad\quad\quad\quad$ **begin** Intersection generation
$\quad\quad\quad\quad\quad$ **for** $i = 1, N$ **do**
$\quad\quad\quad\quad\quad\quad$ **for** $j = 1, N$ **do**
$\quad\quad\quad\quad\quad\quad\quad$ Set $\tau' = \tau$;
$\quad\quad\quad\quad\quad\quad\quad$ Generate $t_{ij}$ trips according to Poisson distribution of parameter $\langle t_{ij} \rangle = fT_d p_{ij} q_h$;
$\quad\quad\quad\quad\quad\quad\quad$ **begin** Trip generation
$\quad\quad\quad\quad\quad\quad\quad\quad$ **for** $t = 1, t_{ij}$ **do**
$\quad\quad\quad\quad\quad\quad\quad\quad\quad$ Generate a time interval $dt$ according to an Exponential distribution of parameter $3600/t_{ij}$;
$\quad\quad\quad\quad\quad\quad\quad\quad\quad$ $\tau' += dt$;
$\quad\quad\quad\quad\quad\quad\quad\quad\quad$ Append $(i, j, t)$ to $\vec{L}$
$\quad\quad\quad\quad\quad\quad\quad\quad$ **end**
$\quad\quad\quad\quad\quad\quad\quad$ **end**
$\quad\quad\quad\quad\quad\quad$ **end**
$\quad\quad\quad\quad\quad$ **end**
$\quad\quad\quad\quad$ **end**
$\quad\quad\quad\quad$ $\tau += 3600$;
$\quad\quad\quad$ **end**
$\quad\quad$ **end**
$\quad$ **end**
**end**
return $\vec{L}$
---

**Algorithm S1. Supersampling algorithm.** Extension of the supersampling algorithm [17] for dynamic allocation in time.

**Supplementary Text**

The shareability network of taxi trips

The shareability curves reported in Figure 1 in the main text are obtained via the notion of *shareability network* defined in [16], which we report here for convenience. Each node of the shareability network represents a trip, and a link between nodes A and B is created if those two trips can be shared (where shareability obeys the rules described in Supplementary Equations). The study [16] is mostly focused on the case where at most two people share a ride. Computing the maximum shareability from a given network thus reduces to the standard problem of computing one of its maximum matchings (a matching is a set of edges without common vertices). We apply that same method to three other cities (San Francisco, Singapore and Vienna) to complement the New York results. As mentioned in the main text, the evolution of shareability with the number of available trips is of particular interest. In a given city and for a given number of trips $N$, we build multiple random subnetworks with $N$ nodes, compute their maximum matching and define the shareability as the average size of those maximum matchings. This gives rise to curves representing the shareability as a function of the number of trips in each city. Since a higher value of $\Delta$ results in a denser shareability network [16], and since that same $\Delta$ is positively correlated to shareability, a positive correlation between shareability and the average node degree of the shareability network is very likely. Conversely, the average node degree is negatively correlated to the fraction of isolated nodes in the shareability network. Hence, one can expect a correlation between shareability $S$ and fraction *iso* of isolated nodes in the network of the type

$$S = a\,(1 - iso) + b.$$

This correlation is consistently observed in the four data sets (Fig. S7) with an $R^2 = 0.991$ for $a = 1.06114$ and $b = -0.0703398$. The strong correlation between shareability and fraction of isolated nodes in the shareability network can be explained as follows. As $\lambda$ increases, a Giant Connected Component (GCC) is rapidly formed in the shareability network, leading to a network composed of a large majority of trips in the GCC, and a few isolated nodes. Since the average node degree in the GCC is very high (e.g., it is about 250 in New York, with $\Delta = 1/12\ h$), its maximum matching is likely to be a perfect matching (a perfect matching is a matching covering all nodes in the network), implying that shareability is well approximated by the fraction of nodes in the GCC. Since the nodes in the shareability network outside the GCC are isolated with overwhelming probability, the fraction of nodes in the GCC (and hence shareability) can be well predicted by the fraction of isolated nodes in the shareability network. This observation is especially important, since it relates shareability -- a quantity determined by the size of the maximum matching in the network -- to one of the network's simplest topological properties, the fraction of isolated nodes. It also suggests that non-shareable trips are likely to occur only at unpopular locations and/or unpopular times.

$R^2$ and rescaling of the shareability curves

The quantity used in this paper to measure similarity between two curves is a classic version of the popular coefficient of determination, known as $R^2$. Given two curves $C_1 := \{(x_1^1, y_1^1),...,(x_n^1, y_n^1)\}$ and $C_2 := \{(x_1^2, y_1^2),...,(x_m^2, y_m^2)\}$, we start by defining a uniformly distributed set of points $\{x_1,..., x_k\}$ in the segment $[max(x_1^1, x_1^2), min(x_n^1, x_m^2)]$. Both curves are then linearly interpolated to obtain two sets $\{(x_1, z_1^1),...,(x_1, z_k^1)\}$ and $\{(x_1, z_1^2),...,(x_1, z_k^2)\}$ representing the values of the curves we want to compare in $x_1,..., x_k$. This allows us to define

$$R^2(C_1, C_2) = 1 - \frac{2}{\sigma_1^2 + \sigma_2^2} \sum_{l=1}^{k}(z_l^1 - z_l^2)^2$$

where $\sigma_i^2$ represents the variance of the set $\{z_k^i,...,z_k^i\}$. The classic $R^2$ usually considers the variance of the data that needs to be calibrated. Here, for the sake of symmetry, we instead average the variance of both curves. Moreover, as long as $k$ is large enough, its choice barely affects the $R^2$ (in this paper, $k = 1000$).

As mentioned throughout this paper, the shareability curves show a very similar shape, and nearly coincide when properly rescaled. The rescaling is applied to the independent variable and hence to the horizontal axis. In other words, if $S(\lambda)$ stands for the shareability corresponding to a given $\lambda$ (the parameter representing the number of trips occurring per hour), the rescaled curve, for a constant scaling factor $K$, is $S_K(\lambda) := S(\lambda / K)$.

For each curve, two rescalings were done. The first is the statistically optimal rescaling, where optimal is defined in the least squares sense: it aims at minimizing the $R^2$ between the curves for two different cities. The values in Table S2 are defined by

$$Opt\ R^2(C_1, C_2) = \inf_{K>0}(R^2(S_K^1(\lambda), S_K^2(\lambda))),$$

where $S^1(\lambda)$ and $S^2(\lambda)$ are the shareability curves for cities $C_1$ and $C_2$. By symmetry of $R^2$, $Opt.\ R^2$ is symmetrical as well. The second rescaling we performed simply corresponds to using $K=L(C)$, as suggested by the dimensional analysis found in the Supplementary Equations. This procedure, which has no adjustable parameters, generates the curve $S_{L(C)}(\lambda)$, plotted for the four cities of interest in Figure 2.

# Supplementary Equations

## 1  General problem

The aim of this paper is to understand the laws governing urban ride sharing and to define a general analytical model capturing their essential features. The model is designed to predict the probability that an arbitrary ride can be shared, given the simple set of urban parameters described below.

Potentially shareable rides, called *trips* in the following, are characterized by their origin, destination and starting time. The model assumes that such a triplet uniquely defines a *trajectory* in the city, and that the average velocity $v$ of trips is specified as a system parameter. Whether any two trips can be shared is determined by spatial and temporal constraints. More specifically, trips $A$ and $B$ can be shared if there exists a route connecting the two origin and destination points such that:

 $a)$  on that route, each origin point precedes its corresponding destination point;

 $b)$  the two trips are overlapped (at least partially), not concatenated; and

 $c)$  the delays imposed on $A$ and $B$ due to sharing are smaller than some tolerable delay $\Delta$, a parameter of the ride sharing system.

Notice that there are exactly four possible routes satisfying conditions $a)$ and $b)$, depending on which trip starts and ends first. The trips are considered shareable if, for one of them at least, condition $c)$ is fulfilled.

The parameters of interest to our study, as well as their qualitative impact on trip shareability, are described below (Table S1 gives their observed values for the taxi systems of New York, San Francisco, Singapore and Vienna):

– *trip density* $\lambda$ (in $trips/h$), which measures the availability of taxis or other potentially shareable rides, and which is expressed as the average number of trips originating in the city per hour. Trip density is both a parameter of the study (through super- and subsampling, see Methods) and a fixed value (denoted $\lambda_f$) when considering the entire datasets of taxi trips. As mentioned in the main text, a clear relationship is found[2] between the trip density and the fraction of trips that can be shared, with a fast increase of the shareability for larger trip density (up to a saturation point).

- *maximum delay* $\Delta$ (in $h$), imposed on passengers sharing their ride. The higher the $\Delta$, the larger the tolerance for shareability, all other parameters being equal. Obviously, a higher shareability is expected for increasing values of $\Delta$.

- *travel speed* $v(C)$ in city $C$ (in $km/h$), assumed for simplicity to be constant in space and time (i.e., in different parts of the city and at different times of day). Travel speed is also positively correlated with shareability: other parameters being equal, higher speed results in the possibility of covering a larger area within the delay bound $\Delta$, and thus yields more sharing opportunities.

- *city area* $|\Omega(C)|$ (in $km^2$), where $\Omega(C)$ denotes the 2D projection of the city. The area is negatively correlated to shareability: all other parameters being equal, a city that is more spread out will offer fewer sharing opportunities than a more compact city would.

## 2 Trip sharing model: overview

Having identified the key parameters influencing trip shareability, we now present a central concept of our mathematical model: the notion of a *shareability shadow*. Assume that the starting times of trips in the city are generated according to a time-dependent Poisson process (Fig. S5). The trips' origins $\mathbf{x}$ and destinations $\mathbf{y}$ are then drawn from a four-dimensional probability distribution, denoted $\rho(\mathbf{x}, \mathbf{y})$. For the sake of simplicity, we assume $\rho$ to be independent of time, but our framework can be generalized to time-varying distributions. At any instant $t$, an existing trip $A$ defines a region of "shareable" origin and destination points:

- Shareable origin points need to be close enough to the current position of $A$.

- Shareable destination points need to be compatible with the destination of $A$, so that $A$ is not forced to deviate too much from its course.

The probability that $A$ can be shared is the probability of a trip to be generated at time $t$ with start and end points in those regions. For the sake of simplicity, the origin and destination regions are assumed independent from one another and of simple shape, as shown in Figs. S2(a) and (b).

We formalize this framework below. It allows us to approximate the probability $\mathcal{S}$ that a trip can be shared, given any spatiotemporal distribution of trip characteristics. The shareability $\mathcal{S}$ takes the form of a five-dimensional (one time dimension, two spatial dimensions for the origin, and two more for the destination) integral of the exponential of another five-dimensional integral. Its analytic evaluation requires choosing a particular form for the spatiotemporal distribution $\rho$. The following section is dedicated to discussing that choice, and the properties we wish our model to encapsulate.

## 3 Fundamental properties and assumptions

As described in the main text and shown in Figure 1, the shareability curves for New York, San Francisco, Singapore and Vienna have extremely similar shapes. After optimal rescaling, the curves lie nearly on top of each other, with $R^2$ values exceeding $95.5\%$; see Figure 2 and Table S2.

The table also shows the similarity between the curves obtained when rescaling $\lambda$ into the dimensionless quantity $L = \lambda v^2 \Delta^3 / |\Omega|$. Although this rescaling is not quite optimal, it has the advantage of having no adjustable parameters, and in that sense can be regarded as universal. As discussed in Section ??, the parameter $L$ accounts in a natural way for the relevant differences between cities. Indeed, putting aside microscopic effects induced by the road network structure, multiplying a city's linear dimensions by a constant $\mu$ (and hence its area by $\mu^2$), while simultaneously multiplying the average speed $v$ in that city by the same constant, keeps $L$ the same and thus should not affect the city's level of shareability.

Likewise, in the interest of prediction and generalization, the spatial distribution for trip generation should be as universal and independent from a city's details as possible. In particular, using an empirically determined function $\rho$ is not necessary to obtain reasonable curves. We show in Section 1.8 that choosing trip origins uniformly in the city, and then destinations uniformly in a disk centered at the origin, generates the required properties, and yields a good fit of the model to the measured shareability curves.

## 4 Notation and definitions

We now describe our mathematical model for trip shareability. Let $\Omega$ be a convex compact subset of $\mathbb{R}^2$, representing the city. *Trips* are uniquely characterized by their origin, destination, and starting time. Formally, any trip $A$ can be expressed as a triplet $(\mathbf{x}_A, \mathbf{y}_A, t_A)$, where $\mathbf{x}_A \in \Omega$ represents the origin of the trip, $\mathbf{y}_A \in \Omega$ its destination and $t_A \in [0, \infty)$ its starting time. Our model assumes that, once $\mathbf{x}_A, \mathbf{y}_A$ and $t_A$ are known, the position of the vehicle and its arrival time $t_A^f$ are fully determined.

In particular, for the sake of mathematical tractability, the trajectory is defined as a straight line joining the origin $\mathbf{x}_A$ to the destination $\mathbf{y}_A$, and the average velocity $v$ of a trip is specified as a system parameter. The arrival time at the destination can be easily estimated as $t_A^f = t_A + ||\mathbf{y}_A - \mathbf{x}_A||/v \equiv t_A + \Delta_A$ where $||\mathbf{x}||$ stands for the Euclidean norm of $\mathbf{x}$ (note that any other distance, e.g. the Manhattan one, could be used here) and $\Delta_A$ for the trip duration. This simple approach allows us to formalize the delay condition for shareability of trips $A$ and $B$: $t_A^{f,S_B} \leq t_A^f + \Delta$ and $t_B^{f,S_A} \leq t_B^f + \Delta$, where $t_A^{f,S_B}$ is the arrival time at destination of trip $A$ when shared with $B$ (similarly for trip $B$), and $\Delta$ is a parameter of the ride sharing system, modeling the maximum delay tolerable to travelers.

Next, we assume that trips $A(\mathbf{x}_A, \mathbf{y}_A, t_A)$ are generated at random in $\Omega$ according to the following rules:

- Starting times $t_A$ are defined as occurrences of a time-dependent Poisson process with rate $\lambda(t)$.

- Origin and destination are chosen according to a two-dimensional probability distribution $\rho(\mathbf{x_A}, \mathbf{y_A})$ independent of the starting time.

For a trip $A(\mathbf{x}_A, \mathbf{y}_A, t_A)$, the position $\mathbf{r}_A(t)$ of the vehicle at time $t$ is entirely determined by the previous hypothesis, and we write it as $\mathbf{r}_A(t) = \mathbf{x}_A + \frac{\mathbf{y}_A - \mathbf{x}_A}{\Delta_A}(t - t_A)$ for any $t \in [t_A, t_A^f]$. We also define the quantity (useful in later derivations):

$$\Gamma_{\varepsilon_1 \frown \varepsilon_2} = \int_{\varepsilon_1} d\mathbf{x}' \int_{\varepsilon_2} d\mathbf{y}' \rho(\mathbf{x}', \mathbf{y}'), \qquad (1)$$

which is the probability a trip generated at a random time $t$ has its origin in $\varepsilon_1 \subseteq \Omega$ and destination in $\varepsilon_2 \subseteq \Omega$ (from now on, all the parameters of the system such as $v$ will be omitted from the notations).

## 5 Probability that a random trip can be shared

Let us consider a trip $A$, starting at time $t_A$. Once its origin $\mathbf{x}_A$ and destination $\mathbf{y}_A$ are chosen, its duration $\Delta_A$ is set as well. Three cases determine the shareability of that trip:

- either it is "backward" shareable because it can be paired with an earlier trip, an event whose probability is written $\mathcal{S}_b(\mathbf{x}_A, \mathbf{y}_A, t_A) = 1 - \mathcal{NS}_b(\mathbf{x}_A, \mathbf{y}_A, t_A)$;

- or it cannot be shared with past trips, in which case it might be "forward" paired with future trips generated in the interval $[t_A, t_A + \Delta_A]$. We let $\mathcal{S}_f(\mathbf{x}_A, \mathbf{y}_A, t_A) = 1 - \mathcal{NS}_f(\mathbf{x}_A, \mathbf{y}_A, t_A)$ denote the probability of that event;

- or it cannot be shared at all.

Thus the probability $\mathcal{S}_b(t)$ for a random trip starting at time $t$ to be shareable due to prior trips (or identically with the subscript $f$ for future trips) is

$$\mathcal{S}_b(t) = \int_{\Omega^2} \mathcal{S}_b(\mathbf{x}, \mathbf{y}, t) \rho(\mathbf{x}, \mathbf{y}) d\mathbf{x} d\mathbf{y} = 1 - \int_{\Omega^2} \mathcal{NS}_b(\mathbf{x}, \mathbf{y}, t) \rho(\mathbf{x}, \mathbf{y}) d\mathbf{x} d\mathbf{y}.$$

And since the events of the random trip being shareable with prior or later trips are independent, the total probability becomes

$$\mathcal{S}(t) = \mathcal{S}_b(t) + (1 - \mathcal{S}_b(t))\mathcal{S}_f(t) = 1 - \mathcal{NS}_b(t)\mathcal{NS}_f(t),$$

where the notations $\mathcal{NS}_b(t)$ and $\mathcal{NS}_f(t)$ are self-explanatory.

## 5.1 Forward probability $\mathcal{NS}_f$

During the time interval $[t_A, t_A^f]$ corresponding to trip $A$, the number $N$ of generated trips follows an exponential distribution with parameter given by $\Lambda(t_A, \Delta_A) = \int_{t_A}^{t_A + \Delta_A} \lambda(t) dt$ (this stems directly from the definition of a Poisson distribution).

Let $p(\mathbf{x}_A, \mathbf{y}_A, t_A)$ denote the probability that trip $A$ can be shared with a random trip generated during $[t_A, t_A^f]$. The probability of $A$ not to be shareable with any of those $N$ trips is $(1 - p(\mathbf{x}_A, \mathbf{y}_A, t_A))^N$, so the average probability $\mathcal{NS}_f$ of trip $A$ not being shared is therefore

$$\mathcal{NS}_f(t_A) = \int_\Omega \int_\Omega d\mathbf{x}_A d\mathbf{y}_A \rho(\mathbf{x}_A, \mathbf{y}_A) \sum_{N \geq 0} e^{-\Lambda} \frac{\Lambda^N}{N!} (1 - p(\mathbf{x}_A, \mathbf{y}_A, t_A))^N.$$

One can rewrite the previous expression to reach the final equation of the formal problem:

$$\mathcal{NS}_f(t_A) = \int_\Omega \int_\Omega d\mathbf{x}_A d\mathbf{y}_A \rho(\mathbf{x}_A, \mathbf{y}_A) \exp\{-\Lambda(t_A, \Delta_A) p(\mathbf{x}_A, \mathbf{y}_A, t_A),\} \equiv \left\langle e^{-f(\mathbf{x}_A, \mathbf{y}_A, t_A)} \right\rangle \quad (2)$$

where we have defined $f(\mathbf{x}_A, \mathbf{y}_A, t_A) \equiv \Lambda(t_A, \Delta_A) p(\mathbf{x}_A, \mathbf{y}_A, t_A)$. We observe that the integral takes the form of the expected value of a negative exponential, and its argument is the average number of trips generated during the duration of $A$ that can be shared with $A$. Note also that, as expected, $0 \leq \mathcal{NS}_f \leq 1$.

The following elements of the problem can be observed in equation (2):

- Duration $\Delta_A = ||\mathbf{y}_A - \mathbf{x}_A||/v$ of trip $A$, and, consequently, trip length $||\mathbf{y}_A - \mathbf{x}_A||$.

- Average number of trips generated during trip $A$, $\Lambda(t_A, \Delta_A) = \int_{t_A}^{t_A+\Delta_A} \lambda(t)dt$.

- Probability $\rho(\mathbf{x}, \mathbf{y})$ of a trip to go from point $\mathbf{x}$ to point $\mathbf{y}$.

- Probability of a trip starting in the interval $[t_A, t_A + \Delta_A]$ to be shareable with $A$, $p(\mathbf{x}_A, \mathbf{y}_A, t_A)$.

These elements reflect the assumptions we have made:

- Trips are generated throughout the city according to a time-dependent Poisson process with rate $\lambda(t)$.

- The spatial generation of trips is characterized by a two-dimensional distribution $\rho(\mathbf{x}, \mathbf{y})$.

- Any trip with given start and end points defines a deterministic trajectory.

Let us now further develop $p(\mathbf{x}_A, \mathbf{y}_A, t_A)$ by taking into account the Poisson nature of the trip generation process. Conditioning on $N$ trips being generated, those $N$ trips are independent events, with starting times distributed in $[t_A, t_A^f]$ according to the probability density function $\lambda(t)/\Lambda(t_A, \Delta_A)$. This yields

$$p(\mathbf{x}_A, \mathbf{y}_A, t_A) = \int_{t_A}^{t_A+\Delta_A} \frac{\lambda(t)}{\Lambda(t_A, \Delta_A)} \mathcal{A}(t|t_A, \mathbf{x}_A, \mathbf{y}_A) dt, \qquad (3)$$

with $\mathcal{A}$ the probability of a trip $X(\mathbf{x}, \mathbf{y}, t)$ generated at time $t \in [t_A, t_A + \Delta_A]$ to be shareable with $A$. For that to happen, two conditions, represented on Fig. S2(a), must be satisfied:

- The starting point of $X$ must be close enough to $\mathbf{r}_A(t)$, which represents the position of vehicle $A$ on its trajectory at time $t$. How close to $\mathbf{r}_A(t)$ that $X$ should be depends on $\Delta$, the delay tolerance of the sharing system, as indicated by $\mathbf{x} \in \varepsilon^\Delta(\mathbf{r}_A(t))$.

- The endpoint of $X$ must be compatible with the trajectory of $A$. This can mean two things. Either $X$ ends before $A$, in which case $\mathbf{y}$ must belong to a region joining $\mathbf{r}_A(t)$ to $\mathbf{y}_A$; if so, let $\Xi(\mathbf{r}_A(t), \mathbf{y}_A)$ denote that region. Otherwise, $X$ ends after $A$, in which case $\mathbf{y}$ must belong to a region surrounding the extension of $A$'s trajectory towards the city's boundaries. We let $\mathbf{x}_B$ denote the intersection of $A$'s trajectory with the boundary. In that case the region can be written as $\Xi(\mathbf{y}_A, \mathbf{x}_B)$.

Those regions form what we call a *shareability shadow* (Fig. S1). Fig. S2(a) shows the shape we chose to give them (properly defined in Section 1.8). Moreover, $\Xi(\mathbf{r}_A(t), \mathbf{y}_A)$ and $\Xi(\mathbf{y}_A, \mathbf{x}_B)$ are disjoint, so the events $\mathbf{y} \in \Xi(\mathbf{r}_A(t), \mathbf{y}_A)$ and $\mathbf{y} \in \Xi(\mathbf{y}_A, \mathbf{x}_B)$ are exclusive. Using the notation defined in (1), the probability $\mathcal{A}$ of $X$ starting in $\varepsilon^\Delta(\mathbf{r}_A(t))$ and ending in $\Xi(\mathbf{r}_A(t), \mathbf{y}_A)$ or $\Xi(\mathbf{y}_A, \mathbf{x}_B)$ can be written as

$$\mathcal{A}(t|t_A, \mathbf{x}_A, \mathbf{y}_A) = \Gamma_{\varepsilon^\Delta(\mathbf{r}_A(t)) \cap \Xi(\mathbf{r}_A(t), \mathbf{y}_A)} + \Gamma_{\varepsilon^\Delta(\mathbf{r}_A(t)) \cap \Xi(\mathbf{y}_A, \mathbf{x}_B)} = \Gamma_{\varepsilon^\Delta(\mathbf{r}_A(t)) \cap \Xi(\mathbf{r}_A(t), \mathbf{x}_B)},$$

where $\Xi(\mathbf{r}_A(t), \mathbf{x}_B)$ corresponds to the union of the green areas in Fig. S2(a).

## 5.2 Backward probability $\mathcal{S}_b$

In the backward case, we study the probability trip $A$ can be shared with a trip prior to it. To do so, we use a procedure analogous to the above calculations. Let us define the following quantities:

- $\Delta_B = ||\hat{\mathbf{x}}_B - \mathbf{x}_A||/v$, the time needed to reach the border of the city ($\hat{\mathbf{x}}_B$ being the point opposite to $\mathbf{x}_B$) in a straight line going "backwards" in the direction of trip $A$ (see Fig. S2(b)). Only trips starting in $[t_A - \Delta_B, t_A]$ can potentially be shared with $A$.

- $\Lambda'(t_A, \Delta_B) = \int_{t_A - \Delta_B}^{t_A} \lambda(t) dt$, the average number of trips generated during the time interval $[t_A - \Delta_B, t_A]$.

- $\mathbf{r}_A(t)$, the virtual position on $A$'s trajectory of a vehicle at time $t \in [t_A - \Delta_B, t_A]$.

- the probability of any past trip to be shareable with $A$:

$$p'(\mathbf{x}_A, \mathbf{y}_A, t_A) = \int_{t_A - \Delta_B}^{t_A} \frac{\lambda(t)}{\Lambda'(t_A, \Delta_B)} \mathcal{A}'(t|t_A, \mathbf{x}_A, \mathbf{y}_A) dt,$$

where

$$\mathcal{A}'(t|t_A, \mathbf{x}_A, \mathbf{y}_A) = \Gamma_{\varepsilon^\Delta(\mathbf{r}_A(t)) \cap \Xi(\mathbf{x}_A, \mathbf{y}_A)} + \Gamma_{\varepsilon^\Delta(\mathbf{r}_A(t)) \cap \Xi(\mathbf{y}_A, \mathbf{x}_B)} = \Gamma_{\varepsilon^\Delta(\mathbf{r}_A(t)) \cap \Xi(\mathbf{x}_A, \mathbf{x}_B)}$$

is the probability a random trip $X$ generated at time $t \in [t_A - \Delta_B, t_A]$ can be shared with $A$. Here $\varepsilon^\Delta(\mathbf{r}_A(t))$ is the region surrounding $\mathbf{r}_A(t)$ where $X$ needs to start for it to be shareable with $A$, while $\Xi(\mathbf{x}_A, \mathbf{x}_B)$ is the region where $X$ needs to end. It can be split into two regions, $\Xi(\mathbf{x}_A, y_A)$ (resp. $\Xi(\mathbf{y}_A, \hat{\mathbf{x}}_B)$) standing for $X$ ending before (resp. after) $A$ ends. Fig. S2(b) shows possible shapes for those three regions.

After some algebra, we obtain

$$\mathcal{NS}_b(t_A) = \int_\Omega \int_\Omega d\mathbf{x}_A d\mathbf{y}_A \rho(\mathbf{x}_A, \mathbf{y}_A) \exp\left\{-\Lambda'(t_A, \Delta_B) p'(\mathbf{x}_A, \mathbf{y}_A, t_A)\right\}.$$

The general setting of the model ends here. We now apply the assumptions described in Section 3.

# 6 Uniform trip generation in time

Let us assume uniform trip generation in time. (This approximation is reasonable if $\lambda(t)$ varies on a time scale longer than a typical trip time.) We have $\lambda(t) = \lambda$ and thus $\Lambda(t_A, \Delta_A) = \lambda \Delta_A$ (and similarly, $\Lambda'(t_A, \Delta_B) = \lambda \Delta_B$). Equations (2) and (3) then become

$$\begin{aligned}
\mathcal{NS}_f &= \int_{\Omega^2} \rho(\mathbf{x}_A, \mathbf{y}_A) \exp\left\{-\lambda \int_{t_A}^{t_A+\Delta_A} \Gamma_{\varepsilon^\Delta(\mathbf{r}_A(t)) \cap \Xi(\mathbf{r}_A(t), \mathbf{x}_B)} dt\right\} d\mathbf{x}_A d\mathbf{y}_A \\
\mathcal{NS}_b &= \int_{\Omega^2} \rho(\mathbf{x}_A, \mathbf{y}_A) \exp\left\{-\lambda \int_{t_A-\Delta_B}^{t_A} \Gamma_{\varepsilon^\Delta(\mathbf{r}_A(t)) \cap \Xi(\mathbf{x}_a, \hat{\mathbf{x}}_B)} dt\right\} d\mathbf{x}_A d\mathbf{y}_A.
\end{aligned} \qquad (4)$$

# 7 Bounded uniform and isotropic spatial generation

The second distribution that needs to be set is the spatial generation of trips. As mentioned above, we assume that trips start uniformly in the city, and that their destination is set at random inside a disk centered at the origin. This boils down to assuming that all the trips have a length $0 \leq l \leq R$: $\rho(\mathbf{x}, \mathbf{y}, t) = \frac{1}{\pi R^2 |\Omega|} \mathbb{1}_{\mathbf{y} \in D(\mathbf{x}, R)}$.

One can further simplify the previous expression by using simple shapes for the shareability shadows $\varepsilon^\Delta(\mathbf{r})$ and $\Xi(\mathbf{x}, \mathbf{y})$. Precisely computing the regions with delays smaller than $\Delta$ for both vehicles is heavy computationally and does not add much to the accuracy of the model. So for simplicity we defined the regions as plain disks and rectangles:

$$\varepsilon^\Delta(\mathbf{r}) = D(\mathbf{r}, \Delta v), \qquad\qquad \Xi(\mathbf{x}, \mathbf{y}) = R(\mathbf{x}, \mathbf{y}, \Delta v),$$

where $D(\mathbf{x}, r)$ stands for a 2-dimensional disk of radius $r$ centered at $\mathbf{x}$ and $R(\mathbf{x}, \mathbf{y}, \Delta v)$ for a rectangle of axis $[\mathbf{x}, \mathbf{y}]$ and width $2\Delta v$. We further disregard border effects, and assume that conditions for shareability are the same for all trips, including those starting close to the city edge. Dealing with border effects would imply cumbersome changes in the mathematical derivations, while providing little improvement in terms of model accuracy due to the fact that the overwhelming majority of trips occur far from the city boundaries.

## 7.1 Forward probability $\mathcal{NS}_f$

The assumptions we just described allow us to develop Equations (4) and compute the shareability as follows:

$$\mathcal{A}(t|t_A, \mathbf{x}_A, \mathbf{y}_A) = \Gamma_{\varepsilon^\Delta(\mathbf{r}_A(t)) \cap \Xi(\mathbf{r}_A(t), \mathbf{x}_B)} \approx \pi \frac{(v\Delta)^2}{|\Omega|} \cdot \frac{2Rv\Delta}{\pi R^2}$$

$$\mathcal{NS}_f = \iint_{\Omega^2} \rho(\mathbf{x}_A, \mathbf{y}_A) \exp\left\{-2\lambda \Delta_A \frac{(v\Delta)^3}{R|\Omega|}\right\} d\mathbf{x}_A d\mathbf{y}_A$$

$$= \int_0^R 2\pi r dr \frac{1}{\pi R^2} \exp\left\{-2\lambda \frac{r}{R} \frac{(v\Delta)^3}{v|\Omega|}\right\} = \frac{1}{2L^2}\left[1 - (1+2L)e^{-2L}\right]$$

where

$$L = \frac{\lambda(v\Delta)^3}{v|\Omega|}.$$

### 7.2 Backward probability $\mathcal{NS}_b$

The backward probability can also be approximated as

$$\mathcal{A}'(t|t_A, \mathbf{x}_A, \mathbf{y}_A) = \Gamma_{\varepsilon^\delta(\mathbf{r}_A(t)) \cap \Xi(\mathbf{x}_A, \mathbf{x}_B)} \approx \pi \frac{(v\Delta)^2}{|\Omega|} \cdot \frac{(R - ||\mathbf{r}_A(t) - x_A||)2\Delta v}{\pi R^2} \mathbb{1}_{||\mathbf{r}_A(t) - x_A|| \leq R}$$

which gives

$$\mathcal{NS}_b = \int_0^R 2\pi r dr \frac{1}{\pi R^2} \exp\left\{-\lambda \frac{R^2 - r^2}{v} \frac{(v\Delta)^3}{R^2|\Omega|}\right\}$$

$$= 2\int_0^1 x dx \exp\left\{-L(1-x^2)\right\} = \frac{1}{L}(1 - e^{-L}).$$

By combining the expressions for the forward and backward probabilities, we obtain the main result of our model: the probability of a trip to be shareable is given by

$$\mathcal{S} = 1 - \mathcal{NS}_f \cdot \mathcal{NS}_b = 1 - \frac{1}{2L^3}(1 - e^{-L})(1 - (1+2L)e^{-2L}). \tag{5}$$

In the next two subsections, it will be convenient to consider shareability as a function of $\lambda$ and $v$, we thus write it $\mathcal{S}(\lambda, v)$.

## 8 Interpolation for time generation

A strong assumption made in the previous analysis is the constant rate in the Poisson process. Trip generation rate is indeed highly dependent on the time of the day, as the analysis of the taxi trip data sets clearly shows; see Fig. S6(D). The most basic approach to address this problem is to disregard hourly fluctuations in trip rate generation, and simply set the trip generation rate as the average daily rate. A more accurate approach, which we call *interpolation* in the following, is to divide the day into hour-long bins, and to consider piecewise constant generation rates during each hour (i.e. assume that rates vary slowly compared to trip durations). Formally, let $\{T_0, T_1, \ldots, T_n\}$ be a subsystem of $[0, T]$ such that $\forall i \in \{0, n-1\}, t \in [T_i, T_{i+1}[, \lambda(t) = \lambda_i$ and the $T_i$ are large compared to the average duration of trips. For such a rate curve, the shareability is equal to

$$\mathcal{S}(\{\lambda_i\}, v) = \frac{\sum_{i=1}^{n} \mathcal{S}(\lambda_i, v)\lambda_i T_i}{\sum_{i=1}^{n} \lambda_i T_i}.$$

Shareability is then computed as a weighted average of the hourly shareability values.

The hourly rates for New York, San Francisco, Singapore, and Vienna are plotted on Fig. S6(D). When studying the impact of the daily number of trips on the shareability, we considered a random subsample of said trips, which boils down to multiplying all the $\lambda_i$ by a number $p \in [0, 1]$. Applying the interpolation to our four cities marginally improves the model accuracy (97.7 to 98.9% for New York, 95.1 to 97.7% for San Francisco, 94.9 to 95.0% for Singapore and 91.2 to 91.4% for Vienna). The limited influence of trip generation rate interpolation on the shareability curves is due to the fact that most trips happen during day time, where the variations in $\lambda$ are quite mild. That observation suggests that the predictions of the model should be accurate also for those cities where the exact hourly rates are unknown.

## 9 Second-order effect on vehicle speed

The average speed of cars in cities is a decreasing function of traffic. Having an analytic function for shareability allows us to take that fact into account rather easily. In a shared economy, the decreased number of cars on the road will generate a lower congestion, a higher average speed, and thus a larger shareability (which is obviously an increasing function of $v$ as stated before and seen from Equation (5). Let us expand the framework by defining the following quantities:

- a function $\tilde{v}(\lambda)$ that associates the density of trips to the average speed of cars in the city,

- a number $\mu$ defining the fraction of people willing to share their trips,

- $\tilde{\lambda}(s) = \lambda - \frac{s}{2}\mu\lambda$, the function giving the trip generation rate if a fraction $s$ of the people willing to share do find a matching trip (we limit our study to at most two people sharing the same ride).

This last point assumes that shareable trips follow the same spatiotemporal distribution as original ones. The "second-order" effect mentioned above can now be taken into account to define the actual shareability in the city. For given $\lambda, \mu > 0$ and city area $|\Omega|$, we define

$$\mathcal{F}\colon [0,1] \to [0,1]$$
$$s \mapsto \mathcal{S}(\lambda\mu, \tilde{v}(\tilde{\lambda}(s))).$$

The shareability that will be reached is the unique fixed point $s^*$ of $\mathcal{F}$. Its existence and uniqueness are guaranteed by the following facts:

- $\mathcal{F}$ is increasing ($\mathcal{S}$ is increasing with respect to its second argument, and $\tilde{v}$ and $\tilde{\lambda}$ are both decreasing),

- $\mathcal{F}(0) > 0$ and $\mathcal{F}(1) < 1$.

To obtain empirical results taking that effect into account requires knowing the function $\tilde{v}(\lambda)$. We do not have it at our disposal. However, the increasing pervasiveness of sensors in the urban area will undoubtedly lead to such knowledge, and eventually to better predictions of shareability.